\documentclass[11pt]{article}
\newcommand{\text}{\rm}

\def\({\c c}
\def\|{\'\i }

\begin{document}

\title{\textbf{Bosonization and Duality Aspects in Superfluids and Superconductors }}
\author{N.C. Ribeiro, R.F. Sobreiro, S.P. Sorella \\
%EndAName
\\
\itshape Universidade do Estado do Rio de Janeiro\\
\itshape Instituto de F\|sica\\
\itshape Departamento de F\|sica Te\'{o}rica\\
\itshape Rua S\~{a}o\ Francisco Xavier, 524, 20550-013 Rio de Janeiro, RJ,
Brazil}
\maketitle

\begin{abstract}
The bosonization and duality rules in three-dimensions are applied
to analyze some features of superfluids and superconductors. The
energy of an ensemble of vortices in a superfluid is recovered by
means of a kind of bound which, to some extent, shares similarity
with the Bogomol'nyi bound. In the case of superconductors, after
recasting the partition function in the form of a pure effective
gauge theory, the existence of finite energy vortex solutions is
discussed.
\end{abstract}

\vfill

\section{Introduction}

In the last years many efforts have been undertaken to generalize the
bosonization techniques \cite{2b} to fermionic models in $(2+1)$ dimensions
\cite{result1}-\cite{result7}. Although a complete framework is still
lacking, these efforts have led to a set of interesting and promising
results, successfully applied to describe transport properties of
interacting electronic systems, such as the universal behavior of the Hall
conductance \cite{hc}. The bosonization rules so far obtained are
efficiently derived by means of the path integral \cite{result2}-\cite
{result6}, being summarized as follows.

Let $j_{\mu }=\bar{\psi}\gamma _{\mu }\psi $ be the $U(1)$ conserved current
of a fermionic system in $(2+1)d$. It turns out that $j_{\mu }$ is mapped
into a topologically conserved current $\,j_{\mu }^{T}$ \cite{result1}-\cite
{result7}, according to
\begin{equation}
j_{\mu }=\bar{\psi}\gamma _{\mu }\psi \,\longrightarrow \,j_{\mu
}^{T}=\epsilon _{\mu \nu \lambda }\partial _{\nu }A_{\lambda }\;.
\label{bcurr}
\end{equation}
From this equation we see that the bosonization of the fermionic current is
achieved through the introduction of an Abelian gauge field $A_{\lambda }$.
Eq.$\left( \ref{bcurr}\right) $ generalizes the well known bosonization rule
in $(1+1)d$, as expressed by the equivalence between the massive Thirring
model and the sine-Gordon theory \cite{2b}. The meaning of the equation $%
\left( \ref{bcurr}\right) $ can be stated in a more precise way at the level
of the current correlation functions, namely

\begin{equation}
\langle j_{\mu _{1}}...j_{\mu _{n}}\rangle _{S_{F}}=\langle j_{\mu
_{1}}^{T}...j_{\mu _{n}}^{T}\rangle _{S_{B}[A]}\ ,  \label{cf}
\end{equation}
where $S_{F}$ is the free massive Dirac action
\begin{equation}
S_{F}[\bar{\psi},\psi ]=\int d^{3}x\,\bar{\psi}\left( i\gamma ^{\mu
}\partial \!\!_{\mu }+m\right) \psi \ ,  \label{faction}
\end{equation}
and $S_{B}[A]$ is the corresponding bosonized action, which displays the
important property of being gauge invariant. We underline that, in general,
a closed form for $S_{B}[A]$ is not available, since its evaluation requires
the exact knowledge of the fermionic determinant in $(2+1)d$ for an
arbitrary gauge field configuration. However, relying only on gauge
invariance, one can show \cite{result6,det} that $S_{B}[A]$ consists of the
Chern-Simons action with the addition of an infinite series of terms,
depending on the curvature $\widetilde{F}_{\mu }=\frac{1}{2}\epsilon _{\mu
\nu \lambda }\partial _{\nu }A_{\lambda }$, \textit{i.e.}
\begin{equation}
S_{B}[A]\;=\;\frac{1}{4\pi }\frac{m}{\left| m\right| }\,S_{CS}[A]\;+\;R[%
\widetilde{F}]\;,  \label{ab}
\end{equation}
where $S_{CS}[A]$ is the Chern-Simons action~
\begin{equation}
S_{CS}[A]\;=\;\frac{1}{2}\int d^{3}x\,\epsilon _{\mu \nu \lambda }A_{\mu
}\partial _{\nu }A_{\lambda }\;,  \label{c-s}
\end{equation}
and $R[\widetilde{F}]$ denotes the remainder, higher order contributions in
the curvature $\widetilde{F}_{\mu }$~\cite{result1}-\cite{result6}. The
presence of the Chern-Simons term in the expression $\left( \ref{ab}\right) $
is expected \cite{p} since the fermionic action $\left( \ref{faction}\right)
$ contains the parity breaking mass term $m\bar{\psi}\psi $ .

$\ $It is worth mentioning that the bosonization rule $\left( \ref{cf}%
\right) $ has been proven to hold \cite{un} also in the presence of a large
class of interaction terms $I[j^{\mu }]$ depending only on the fermionic
current, namely

\begin{eqnarray}
S_{F}[{\bar{\psi}},\psi ]+I[j^{\mu }] &\leftrightarrow &S_{B}[A]+I[\epsilon
^{\mu \nu \rho }\partial _{\nu }A_{\rho }]\;,  \label{u} \\
\langle j_{\mu _{1}}...j_{\mu _{n}}\rangle _{(S_{F}+I[j])} &=&\langle j_{\mu
_{1}}^{T}...j_{\mu _{n}}^{T}\rangle _{(S_{B}+I[j^{T}])}\ .  \nonumber
\end{eqnarray}
Notice that the bosonization of the interaction term $I[j^{\mu }]$ is
obtained by replacing the fermionic current $j^{\mu }=$ $\bar{\psi}\gamma
_{\mu }\psi \,$by the topological current $j_{\mu }^{T}=\epsilon _{\mu \nu
\lambda }\partial _{\nu }A_{\lambda }$, without changing the functional form
of the interaction term $I[j^{\mu }]$. This important property gives a
universal character to the bosonization rules $\left( \mathrm{{\ref{u}}}%
\right) $. For instance, in \cite{bi}, the massive Thirring model with a
non-polynomial current-current interaction has been bosonized into the $%
(2+1)d$ Born-Infeld gauge theory, according to eqs.$\left( \mathrm{{\ref{u}}}%
\right) $.

More recently, the authors \cite{dual} have been able to generalize this
construction, originally used for the bosonization of fermionic actions, to
the case of an arbitrary field theory model with a conserved global $U(1)$
charge. It has been proven that, regardless of the spin and statistics of
the fields in the original action, any three-dimensional model with an
Abelian conserved charge can be mapped into a dual Abelian gauge theory \cite
{dual}. This mapping acquires therefore the meaning of a \emph{duality\/},
in view of its more general defining characteristics. As the most
distinctive feature of this mapping, we mention that the conserved Noether
current $j_{\mu }^{U(1)}$ corresponding to the $U(1)$ global symmetry has,
again, a topological current $j_{\mu }^{T}$ as its dual partner in the
Abelian gauge theory, \textit{i.e.}
\begin{equation}
j_{\mu }^{U(1)}\,\longrightarrow \;j_{\mu }^{T}=\epsilon _{\mu \nu \lambda
}\partial _{\nu }A_{\lambda }\;.  \label{u1-curr}
\end{equation}
Equations $\left( \mathrm{{\ref{u}}}\right) $ have also their
generalization, which reads
\begin{equation}
S[\phi ,\phi ^{\dagger }]+I[j_{\mu }^{U(1)}]\leftrightarrow S_{\mathrm{dual}%
}[A]+I[\epsilon _{\mu \nu \lambda }\partial _{\nu }A_{\lambda }]\ ,
\label{udual}
\end{equation}
where $\phi ,\phi ^{\dagger }$ are a pair of generic complex $U(1)$ charged
fields, $S[\phi ,\phi ^{\dagger }]$ is the corresponding invariant action,
and $I[j_{\mu }^{U(1)}]$ collects the interaction terms depending only on
the Noether current.

In the present work we shall apply the dual mapping $\left( \mathrm{{\ref
{u1-curr}}}\right) -\left( \mathrm{{\ref{udual}}}\right) $ to analyze some
features of superfluids and superconductors, the aim being to show that the
dual description in terms of gauge theories can be effectively worked out.
As first example, we shall consider the superfluid $^{4}\mathrm{He}$ system
in the presence of an ensemble of vortices. The dual mapping will be used to
evaluate the energy of the vortices. In particular, we shall be able to
prove that the energy configuration can be obtained by means of a
relationship which, to some extent, shares close similarity with the
Bogomol'nyi bound of superconductors \cite{bg}. In this case the mapping $%
\left( \mathrm{{\ref{u1-curr}}}\right) -\left(
\mathrm{{\ref{udual}}}\right) $ will have the meaning of a duality
between vortices and electric charges. A vortex line can be
regarded in fact as the world-line of a particle whose electric
charge is proportional to the topological integer corresponding to
the so called strength of the vortex. Although the duality between
vortices in $^{4}\mathrm{He}$ and electric charges is already
known \cite{kl,po}, we
point out that it is a natural consequence of the relation $\left( \mathrm{{%
\ref{u1-curr}}}\right) $. In the second example, after recasting
the partition function of the Landau-Ginzburg model for a
superconductor in the form a pure three-dimensional gauge theory,
we shall analyze the existence of vortex configurations in the
resulting effective theory. We remark here that these examples
will be worked out in the Euclidean three-dimensional
space $R^{3}$ rather than in $(2+1)d$. In this case, the bosonization rules $%
\left( \mathrm{{\ref{u}}}\right) $ and the dual mapping $\left( \mathrm{{\ref
{u1-curr}}}\right) -\left( \mathrm{{\ref{udual}}}\right) $ turn out \cite
{dual} to describe the static properties of three-dimensional systems.

It is worth underlining that the possibility of describing
three-dimensional $U(1)$ charged models in terms of the
corresponding dual gauge theory can be useful in order to improve
our understanding of the nonperturbative aspects of these models.
For instance, in the case of superconductors, it turns out that
the resulting effective dual gauge theory is described by an
action containing a term of the type $\int d^{3}x\left(
F_{ij}+J_{ij}\right) ^{2}$, where $J_{ij}$ accounts for the
presence of vortex lines. This kind of action is deeply related to
the formulation of the so called compact $U(1)$ gauge theories in
continuum space-time. These models are object of intensive
investigation in order to obtain a better understanding of the
phase structure of confining theories and in order to test new
order parameters, as in the case of the operator $A_{\mathrm{\min
}}^{2}$ recently proposed in the context of three-dimensional
compact $QED$ \cite{z}. Also, this kind of action naturally arises
in  $SU(2)$ Yang-Mills theories which, after performing 't Hooft
abelian projection, turn out to be described by an effective
$U(1)$ compact theory \cite{an}, which is expected to be the basic
tool for the dual superconductivity mechanism for the confinement
of quarks and gluons.

\section{Vortices in the $^{4}\mathrm{He}$ Superfluid}

\subsection{Generalities}

In this section we shall briefly review some basic features of the $^{4}%
\mathrm{He}$ superfluid in the presence of an ensemble of vortices \cite{kl}%
. Let us begin with the expression of the Landau-Ginzburg free energy for
the $^{4}\mathrm{He}$ in the superfluid phase

\begin{equation}
E=\int d^{3}x\partial _{i}\varphi ^{+}\partial _{i}\varphi +\frac{\lambda }{2%
}\left( \varphi ^{+}\varphi -\sigma ^{2}\right) ^{2}\;,  \label{lg}
\end{equation}
where the complex field $\varphi $ stands for the superfluid order parameter
and the index $i$ runs from $1$ to $3$. For temperatures $T\ll T_{\lambda }$
well below the critical value $T_{\lambda }\simeq 2.18K$ corresponding to
the so called superfluid $\lambda -$transition, only the phase fluctuations
of the order parameter are relevant, namely

\begin{equation}
\varphi (x)=\sigma e^{i\gamma (x)}\;,  \label{ph}
\end{equation}
where the parameter $\sigma $ is related to the density of the superfluid
component \cite{kl}, identified with the Bose-Einstein condensate.
Therefore, for the Landau-Ginzburg free energy we get

\begin{equation}
E=\sigma ^{2}\int d^{3}x\partial _{i}\gamma \partial _{i}\gamma \;.
\label{eg}
\end{equation}
The phase configurations which minimize the functional $\left( \mathrm{{\ref
{eg}}}\right) $ are those obeying the Poisson equation

\begin{equation}
\partial _{i}\partial _{i}\gamma =0\;.  \label{p}
\end{equation}
The presence of vortices in the superfluid is associated with the
existence of lines of singularity in the phase $\gamma $, which
turns out to be multi-valued. These lines can be of any shape and
they never end inside the
superfluid. They can be closed or run to infinity\footnote{%
In this last case they extend from $-\infty $ to $+\infty $, being meant to
be closed at infinity.}. For instance, an ensemble of $N$ vortices
corresponding to a set of closed lines $\left\{ L^{I},\;I=1,...,N\right\} $
is described by

\begin{equation}
\epsilon _{ijk}\partial _{j}\partial _{k}\gamma (x)=\sum_{I}2\pi
n^{(I)}\int_{L^{I}}dy_{i}\delta ^{3}(x-y)\;,  \label{vc}
\end{equation}
where $L^{I}$ denotes a smooth closed curve and $n^{(I)}$ is an integer.
From this expression it is manifest that the phase $\gamma $ is singular
along each curve $L^{I}$, the singularity being expressed by the line
integral of the $\delta $-function, \textit{i.e. }$\int_{L^{I}}dy_{i}\delta
^{3}(x-y)$. Since the curves $\left\{ L^{I}\right\} $ are closed, it follows
that

\begin{equation}
\epsilon _{ijk}\partial _{i}\partial _{j}\partial _{k}\gamma (x)=0\;.
\label{cl}
\end{equation}
Also, from eq.$\left( \mathrm{{\ref{vc}}}\right) $, we see that to each
vortex line $L^{I}$ there corresponds an integer $n^{(I)}$, known as the
strength of the vortex. Before ending this short summary, it is useful to
mention that the vortices have a finite size, looking alike thin tubes of
radius $\xi $, where $\xi $ is the coherence length. The region $r\leq \xi $
is called the core of the vortex. The superfluid does not penetrate inside
the core region, which contributes to total energy of the fluid. For an
estimate of the core energy see \cite{kl}.

\section{Evaluation of the energy of an ensemble of vortices}

Let us turn now to the evaluation of the energy $E$ in the expression $%
\left( \mathrm{{\ref{eg}}}\right) $ for the ensemble of vortices specified
by the eq.$\left( \mathrm{{\ref{vc}}}\right) $. In order to make use of the
dual mapping $\left( \mathrm{{\ref{u1-curr}}}\right) -\left( \mathrm{{\ref
{udual}}}\right) $, we begin by computing the $U(1)$ current $j_{i}^{U(1)}$
for  the Landau-Ginzburg functional $\left( \mathrm{{\ref{lg}}}\right) $. We
easily find

\begin{equation}
j_{i}^{U(1)}=\frac{i}{2}\left( \left( \partial _{i}\varphi ^{+}\right)
\varphi -\left( \partial _{i}\varphi \right) \varphi ^{+}\right) =\sigma
^{2}\partial _{i}\gamma \;\mathrm{.}  \label{ch}
\end{equation}
According to eqs.$\left( \mathrm{{\ref{u1-curr}}}\right) $, $\left( \mathrm{{%
\ref{udual}}}\right) $, the free energy $\left( \mathrm{{\ref{eg}}}\right) $
has a dual description in terms of a gauge theory, the dual gauge field $%
A_{i}$ being identified through the topological current $j_{i}^{T}=\epsilon
_{ijk}\partial _{j}A_{k}$, namely
\begin{equation}
\sigma ^{2}\partial _{i}\gamma =\epsilon _{ijk}\partial _{j}A_{k}\;.
\label{gf}
\end{equation}
As it will become clear in the following, it is useful to introduce the
quantity $\Theta $ defined by

\begin{equation}
\Theta =\int d^{3}x\left( \sigma ^{2}\partial _{i}\gamma -\epsilon
_{ijk}\partial _{j}A_{k}\right) ^{2}\;,  \label{th}
\end{equation}
where, for the time being, $A_{i}$ denotes a generic field configuration,
not yet identified with the dual gauge field of the eq.$\left( \mathrm{{\ref
{gf}}}\right) $. Obviously, we have

\begin{equation}
\Theta \geq 0\;,  \label{bb}
\end{equation}
the equality holding when $A_{i}$ is the dual field. The condition $\left(
\mathrm{{\ref{bb}}}\right) $ can be seen as a kind of bound which, to some
extent, shares similarity with the Bogomol'nyi bound of superconductors \cite
{bg}. The bound is saturated when $\Theta =0$, \textit{i.e. }when the gauge
field is identified with the dual field of eq.$\left( \mathrm{{\ref{gf}}}%
\right) $. Notice also that, as in the case of the Bogomol'nyi bound, the
expression $\left( \mathrm{{\ref{gf}}}\right) $ is of first order in the
derivatives of the fields.

Expression $\left( \mathrm{{\ref{th}}}\right) $ is easily worked out,
yielding

\begin{equation}
\Theta =\int d^{3}x\left( \sigma ^{4}\partial _{i}\gamma \partial _{i}\gamma
+\frac{1}{2}F_{ij}F_{ij}-2\sigma ^{2}\epsilon _{ijk}A_{i}\partial
_{j}\partial _{k}\gamma \right) \;.  \label{thex}
\end{equation}
Requiring now that $A_{i}$ is precisely the dual gauge field of eq.$\left(
\mathrm{{\ref{gf}}}\right) $, the bound gets saturated so that $\Theta =0$.
Thus, for the energy of the ensemble of vortices we get

\begin{equation}
E=\sigma ^{2}\int d^{3}x\partial _{i}\gamma \partial _{i}\gamma \;=\int
d^{3}x\left( -\frac{1}{2\sigma ^{2}}F_{ij}F_{ij}+2A_{i}I_{i}\right) \;,
\label{bs}
\end{equation}
with $F_{ij}=\left( \partial _{i}A_{j}-\partial _{j}A_{i}\right) $ and
\begin{equation}
I_{i}(x)=\sum_{I}2\pi n^{(I)}\int_{L^{I}}dy_{i}\delta ^{3}(x-y)\;.
\label{ci}
\end{equation}
The equation $\left( \mathrm{{\ref{bs}}}\right) $ shows that, according to
the dual mapping $\left( \mathrm{{\ref{u1-curr}}}\right) -\left( \mathrm{{%
\ref{udual}}}\right) $, the energy of an ensemble of vortex lines can be
obtained by evaluating the action of an electromagnetic field interacting
with the external current $I_{i}(x)$. Furthermore, we notice that the
current $I_{i}(x)$ in eq.$\left( \mathrm{{\ref{ci}}}\right) $ corresponds to
a system of charged particles whose electric charges are given by $%
q^{(I)}=4\pi n^{(I)}$. In addition, the vortex line integrals $%
\int_{L^{I}}dy_{i}\delta ^{3}(x-y)$ can be associated to the
world-lines of
the particles, making manifest the duality aspects between vortices in the $%
^{4}\mathrm{He}$ superfluid and a system of electric charges. We also remark
that, from the Poisson equation $\left( \mathrm{{\ref{p}}}\right) $, it
follows that

\begin{equation}
\epsilon _{ijk}\partial _{i}\partial _{j}A_{k}=0\;,  \label{bid}
\end{equation}
which implies the absence of magnetic charges.

In order to evaluate the electromagnetic action in eq.$\left( \mathrm{{\ref
{bs}}}\right) $, we act with the operator $\epsilon _{mni}\partial _{n}$ on
both sides of the equation $\left( \mathrm{{\ref{gf}}}\right) $, obtaining
\begin{equation}
\partial _{k}\partial _{k}A_{i}=-\sigma ^{2}I_{i}\;,  \label{d2}
\end{equation}
where use has been made of the gauge condition $\partial _{k}A_{k}=0$. For
the dual gauge field we get
\begin{equation}
A_{i}(x)=\frac{\sigma ^{2}}{4\pi }\int d^{3}y\frac{1}{|\overrightarrow{x}-%
\overrightarrow{y}|}I_{i}(y)\;.  \label{sp}
\end{equation}
Thus,
\begin{equation}
\int d^{3}x\left( -\frac{1}{2\sigma ^{2}}F_{ij}F_{ij}+2A_{i}I_{i}\right) =%
\frac{\sigma ^{2}}{4\pi }\int d^{3}\overrightarrow{y}d^{3}\overrightarrow{x}%
I_{i}\left( \overrightarrow{y}\right) \frac{1}{|\overrightarrow{x}-%
\overrightarrow{y}|}I_{i}\left( \overrightarrow{x}\right) \;.  \label{ea}
\end{equation}
Finally, the energy of the ensemble of vortices if found to be
\begin{equation}
E=\sigma ^{2}\int d^{3}x\partial _{i}\gamma \partial _{i}\gamma =\sigma
^{2}\pi \sum_{I,J}n^{(I)}n^{(J)}\int_{L^{I}}d\overrightarrow{x}^{I}\cdot
\int_{L^{J}}d\overrightarrow{y}^{J}\frac{1}{\left| \overrightarrow{x}^{I}-%
\overrightarrow{y}^{I}\right| }\;,  \label{efv}
\end{equation}
which coincides in fact with the well known expression given in \cite{kl,po}%
. We have thus recovered the energy of an ensembles of vortex lines by using
the dual mapping.

\section{Effective gauge theory for superconductors}

In this second example we shall consider the partition function of a
superconductor. After recasting it in the form of a pure gauge theory, we
shall discuss the existence of finite energy vortex solutions in the
resulting effective theory.

The partition function of the Landau-Ginzburg free energy of a
superconductor reads
\begin{equation}
\mathcal{Z=}\int DA_{i}D\varphi ^{+}D\varphi \exp \left[ -\int d^{3}x\left(
\frac{1}{4}F_{ij}F_{ij}+\left( \mathcal{D}_{i}{\varphi }\right) ^{+}\left(
\mathcal{D}_{i}{\varphi }\right) +V(\varphi )\;\right) \right] \;,
\label{lgs}
\end{equation}
where $\varphi $ stands now for the order parameter describing the
condensation of the Cooper pairs and
\begin{equation}
V{(\varphi )=}\frac{\lambda }{2}\left( \varphi ^{+}\varphi -\sigma
^{2}\right) ^{2}\;.  \label{ps}
\end{equation}
The covariant derivative is defined as
\[
\mathcal{D}_{i}=\partial _{i}-iqA_{i}
\]
with $q=2e$ the charge of a Cooper pair. In order to obtain a representation
of the partition function $\left( \mathrm{{\ref{lgs}}}\right) $ in terms of
an effective gauge theory, we first write the order parameter $\varphi $ as

\begin{equation}
\varphi =(\sigma +\rho )e^{i\theta (x)}\;,  \label{op}
\end{equation}
where $\rho $ accounts for the fluctuations above the condensate
$\sigma $. Further, we decompose the phase $\theta $ according to

\begin{equation}
\theta =\theta ^{\mathrm{r}}+\theta ^{\mathrm{s}}\;,  \label{d}
\end{equation}
where $\theta ^{\mathrm{r}}$, $\theta ^{\mathrm{s}}$ denote the regular and
the singular part of $\theta $ \cite{ant}, namely

\begin{eqnarray}
\epsilon _{ijk}\partial _{j}\partial _{k}\theta ^{\mathrm{r}}(x) &=&0\;,
\nonumber \\
\epsilon _{ijk}\partial _{j}\partial _{k}\theta ^{\mathrm{s}}(x)
&=&I_{i}(x)=2\pi n\int_{L}dy_{i}\delta ^{3}(x-y)\;,  \label{rs}
\end{eqnarray}
where $n$ is an integer. As in the case of the $^{4}$\textrm{He} superfluid,
the quantity $I_{i}(x)$ specifies the singularity of $\theta ^{\mathrm{s}%
}(x) $, which turns out to correspond to Abrikosov-Nielsen-Olesen (ANO) \cite
{ano} flux tubes inside the superconductor. We shall limit here to consider
the case of a unique smooth curve$\;L$, the example being easily generalized
to an ensemble of curves.

Making use of the change of variables

\begin{eqnarray}
A_{i} &\rightarrow &A_{i}+\frac{1}{q}(\partial _{i}\theta ^{\mathrm{r}%
}+\partial _{i}\theta ^{\mathrm{s}})\;,  \label{cv} \\
\theta ^{\mathrm{r}} &\rightarrow &\theta ^{\mathrm{r}}\;\;,\;\theta ^{%
\mathrm{s}}\rightarrow \theta ^{\mathrm{s}}\;,\;\rho \rightarrow \rho \;,
\nonumber
\end{eqnarray}
and noticing that, due to eq.$\left( \mathrm{{\ref{rs}}}\right) $,
\[
F_{ij}\rightarrow F_{ij}+J_{ij},\hspace{0.25in}J_{ij}=\frac{1}{q}\epsilon
_{ijk}I_{k}
\]
for the partition function we get

\begin{equation}
\mathcal{Z}=\mathcal{N\;}\int DA_{i}D\theta ^{\mathrm{s}}e^{-E_{\mathrm{eff}%
}(A,J)}\;,  \label{pf}
\end{equation}
where $\mathcal{N}$ is a normalization factor and $E_{\mathrm{eff}}(A,J)$
stands for the effective energy functional corresponding to

\begin{equation}
E_{\mathrm{eff}}(A,J)=\int d^{3}x\left( \frac{1}{4}\left(
F_{ij}+J_{ij}\right) ^{2}+\sigma ^{2}q^{2}A^{2}\right) +\widetilde{E}(A)\;,
\label{effg}
\end{equation}
$\widetilde{E}(A)$ denoting the contribution coming from the integration
over the radial fluctuations $\rho $, namely

\begin{equation}
e^{-\widetilde{E}(A)}=\int D\rho \exp \left[ -\int d^{3}x\left( \partial
_{i}\rho \partial _{i}\rho +q^{2}\rho (\rho +2\sigma )A^{2}+V(\rho )\right)
\right] \;.  \label{pr}
\end{equation}
Although we shall be interested in the study of the classical vortex
solutions of $E_{\mathrm{eff}}(A,J)$, it is worth mentioning that the
integration over the singular variable $D\theta ^{\mathrm{s}}$ in the
expression $\left( \mathrm{{\ref{pf}}}\right) $ can be specified in a more
precise way as an integration over all possible smooth curves $\left\{
L\right\} $, as discussed in \cite{kl,ant}.

The expression $\left( \mathrm{{\ref{pf}}}\right) $ yields the
representation of the partition function $\mathcal{Z}$ in terms of the
effective gauge theory we were looking for. The resulting effective
functional $E_{\mathrm{eff}}(A,J)$ in eq.$\left( \mathrm{{\ref{effg}}}%
\right) $ is sometimes referred as to a compact gauge theory, due to the
presence of singular configurations, encoded in the line integral $J_{ij}$
and in the functional measure $D\theta ^{\mathrm{s}}$\cite{kl,ant}.

Let us turn now to the study of the existence of finite energy vortex
configurations for the resulting energy functional $E_{\mathrm{eff}}(A,J)$.
We shall begin by working out the so called London limit, corresponding to $%
\lambda \rightarrow \infty $.

\subsection{Vortex solution in the London limit}

In the London limit, \textit{i.e. }$\lambda \rightarrow \infty $, the radial
part $\rho $ of the order parameter gets frozen

\begin{equation}
\varphi =\sigma e^{i\theta (x)}\;.  \label{fr}
\end{equation}
As a consequence $\widetilde{E}(A)=0$, so that $E_{\mathrm{eff}}(A,J)$
reduces to

\begin{equation}
E_{\mathrm{Lond}}(A,J)=\int d^{3}x\left( \frac{1}{4}\left(
F_{ij}+J_{ij}\right) ^{2}+\frac{1}{2}m^{2}A_{i}A_{i}\right) \;\;\;,\;\;\;\;
\label{l}
\end{equation}
with
\begin{equation}
m^{2}=2\sigma ^{2}q^{2}\;.  \label{m}
\end{equation}
The configurations which minimize the functional $E_{\mathrm{Lond}}(A,J)$
are those for which

\begin{equation}
\frac{\delta E_{\mathrm{Lond}}}{\delta A_{i}}=-\partial
_{i}F_{ij}+m^{2}A_{j}-\partial _{i}J_{ij}=0\;.  \label{eqm}
\end{equation}
From eq.$\left( \mathrm{{\ref{eqm}}}\right) $ it follows that $A_{i}$ is
transverse
\begin{equation}
\partial _{i}A_{i}=0\;.  \label{tr}
\end{equation}
Therefore, expression $\left( \mathrm{{\ref{eqm}}}\right) $ becomes

\begin{equation}
(\partial ^{2}-m^{2})A_{i}=\frac{1}{q}\epsilon _{ijk}\partial _{j}I_{k}\;.
\label{bc}
\end{equation}
From

\begin{eqnarray}
(-\partial ^{2}+m^{2})\mathcal{G}(x-y) &=&\delta ^{3}(x-y)\;,  \label{grn} \\
\mathcal{G}(x-y) &=&\frac{1}{(-\partial ^{2}+m^{2})}=\frac{1}{4\pi }\frac{1}{%
\left| x-y\right| }e^{-m\left| x-y\right| }\;,  \nonumber
\end{eqnarray}
we get

\begin{equation}
A_{i}(x)=-\frac{1}{4\pi }\frac{1}{q}\epsilon _{ijk}\partial _{j}^{x}\int
d^{3}y\frac{1}{\left| x-y\right| }e^{-m\left| x-y\right| }I_{k}(y)\;,
\label{ai}
\end{equation}
which, from eq.$\left( \mathrm{{\ref{rs}}}\right) $, takes the final form

\begin{equation}
A_{i}(x)=-\frac{n}{2q}\epsilon _{ijk}\partial _{j}^{x}\int_{L}dy_{k}\frac{1}{%
\left| x-y\right| }e^{-m\left| x-y\right| }\;.  \label{aif}
\end{equation}
This expression yields the solution of the equations of motion for a generic
smooth curve $L$. It corresponds in fact to a vortex solution localized
around the curve $L$, on a scale of the order $1/m$. For a better
understanding of expression $\left( \mathrm{{\ref{aif}}}\right) $ let us
work out in details the case in which $L$ is a straight line coinciding with
the $z$-axis. In this case, the line integral can be evaluated in closed
form, giving

\begin{eqnarray}
A_{i} &=&-\frac{n}{q}\epsilon _{ij3}\partial _{j}K_{0}(mx_{\mathrm{\bot }%
})\;,  \label{az} \\
A_{x} &=&-\frac{n}{q}\partial _{y}K_{0}(mx_{\mathrm{\bot }})\;,\;\;\;\;A_{y}=%
\frac{n}{q}\partial _{x}K_{0}(mx_{\mathrm{\bot }})\;,\;\;\;A_{z}=0\;,
\nonumber
\end{eqnarray}
where $K_{0}$ is the Bessel function and $x_{\mathrm{\bot }}=\sqrt{%
x^{2}+y^{2}}$ is the distance from the origin in the $\left( x,y\right) -$%
plane. The magnetic field $B=B_{z}$ is found

\begin{equation}
B=\partial _{x}A_{y}-\partial _{y}A_{x}=\frac{n}{q}\frac{1}{x_{\mathrm{\bot }%
}}\partial _{x_{\mathrm{\bot }}}\left( x_{\mathrm{\bot }}\partial _{x_{%
\mathrm{\bot }}}K_{0}(mx_{\mathrm{\bot }})\right) \;.  \label{b}
\end{equation}
Also, for the flux of $B$ through the circle at infinity in the $\left(
x,y\right) -$plane, one obtains
\begin{equation}
\Phi =\int_{\mathrm{S}_{\infty }^{1}}d^{2}xB=\frac{2\pi n}{q}%
\int_{0}^{\infty }dx_{\mathrm{\bot }}\partial _{x_{\mathrm{\bot }}}\left( x_{%
\mathrm{\bot }}\partial _{x_{\mathrm{\bot }}}K_{0}(mx_{\mathrm{\bot }%
})\right) =\frac{2\pi n}{q}\;.  \label{fl}
\end{equation}
The gauge configuration $\left( \mathrm{{\ref{az}}}\right) $ is nothing but
the London limit of the (ANO) vortex solution. In particular, the equation $%
\left( \mathrm{{\ref{fl}}}\right) $ expresses the well known flux
quantization of the ANO vortex in superconductors. We have thus checked that
the effective gauge functional $E_{\mathrm{Lond}}(A,J)$ of eq.$\left(
\mathrm{{\ref{l}}}\right) $ admits a vortex solution with the correct
quantization of the magnetic flux. Moreover, because of the limit $\lambda
\rightarrow \infty $, the configuration $\left( \mathrm{{\ref{az}}}\right) $
does not fulfill the requirement of finite energy. As we shall discuss in
the next section, this feature can be handled by going beyond the London
limit.

\subsection{Finite energy solutions}

So far we have seen that the functional $E_{\mathrm{Lond}}(A,J)$ of eq.$%
\left( \mathrm{{\ref{l}}}\right) $ admits vortex configurations, although
they do not meet the requirement of finite energy. This means that the
classical configuration $\left( \mathrm{{\ref{az}}}\right) $ gives a
divergent value for $E_{\mathrm{Lond}}(A,J)$. Indeed, from eq.$\left(
\mathrm{{\ref{l}}}\right) $ we have

\begin{equation}
E_{\mathrm{Lond}}(A,J)=\int d^{3}x\left( \frac{1}{2}A_{i}\left( -\partial
^{2}+m^{2}\right) A_{i}+\frac{1}{q}A_{i}\epsilon _{ijk}\partial _{j}I_{k}+%
\frac{1}{2q^{2}}I_{k}I_{k}\right) \;.  \label{ld}
\end{equation}
From

\begin{equation}
A_{i}=-\frac{1}{q}\frac{\epsilon _{ijk}}{\left( -\partial ^{2}+m^{2}\right) }%
\partial _{j}I_{k}\;,  \label{ldv}
\end{equation}
one easily gets
\begin{equation}
E_{\mathrm{Lond}}(A,J)=\frac{m^{2}}{2q^{2}}\int d^{3}xI_{k}\frac{1}{\left(
-\partial ^{2}+m^{2}\right) }I_{k}\;.  \label{ldd}
\end{equation}
In the case in which $I_{k}$ is a straight line coinciding with the $z$%
-axis, expression $\left( \mathrm{{\ref{ldd}}}\right) $ is easily written in
momentum space, yielding
\begin{equation}
\frac{E_{\mathrm{Lond}}}{L}=\frac{m^{2}n^{2}}{2q^{2}}\int d^{2}k_{\bot }%
\frac{1}{k_{\bot }^{2}+m^{2}}\;,  \label{luv}
\end{equation}
where $E_{\mathrm{Lond}}/L$ is the energy per unit length$\footnote{%
As usual, we have factorized the integration over the $z$-axis,\textit{\
i.e. } $\int dz=L$, due to the independence of the integrand in eq.$\left(
\mathrm{{\ref{ldd}}}\right) $ from the $z$ coordinate.}$ and $%
\overrightarrow{k}_{\bot }$ is the two-dimensional vector $\overrightarrow{k}%
_{\bot }=(k_{x},k_{y},0)$. We see that $E_{\mathrm{Lond}}/L$ displays an
ultraviolet logarithmic divergence in momentum space. We need therefore to
go beyond the London limit, which amounts to evaluate the quantity $%
\widetilde{E}(A)$ in expression $\left( \mathrm{{\ref{effg}}}\right) $. This
is a rather difficult task, since it requires the evaluation of the
functional integral of eq.$\left( \mathrm{{\ref{pr}}}\right) $, whose exact
form is unknown, even if the cubic and quartic terms contained in the
potential $V(\rho )$ are neglected\footnote{%
See \cite{bl} for a recent attempt to go beyond the London limit in the case
of the four-dimensional euclidean dual Abelian Higgs model.}. Nevertheless,
due to the gauge invariance, we expect that $\widetilde{E}(A)$ will contain
nonlocal terms in the curvature $F_{ij}$. It is worth to point out here that
these nonlocal terms naturally arise in the evaluation of three-dimensional
effective action as, for instance, in the case of the bosonized action
corresponding to the fermionic determinant \cite{result6,det}. Moreover, as
shown in \cite{os}, these nonlocal terms provide a useful ultraviolet
regularization, allowing in fact to fulfill the requirement of finite
energy. Following \cite{os}, a simple understanding of the regularizing
mechanism due to the presence of nonlocal terms can be achieved by replacing
in the effective functional $E_{\mathrm{Lond}}(A,J)$ of eq.$\left( \mathrm{{%
\ref{ld}}}\right) $ the field strength $F_{ij}$ by
\begin{equation}
F_{ij}\rightarrow \widehat{O}(\partial )F_{ij}\;,  \label{rf}
\end{equation}
where $\widehat{O}$ is a suitable nonlocal operator associated with a kernel
$O(x-y)$, according to
\begin{equation}
\left[ \widehat{O}f\right] (x)=\int d^{3}yO(x-y)f(y)\;.  \label{odf}
\end{equation}
With the replacement $\left( \mathrm{{\ref{rf}}}\right) $, the functional $%
E_{\mathrm{Lond}}(A,J)$ becomes
\begin{equation}
E_{\mathrm{Lond}}(A,J)\rightarrow E_{O}(A,J)\;,  \label{ereg}
\end{equation}
with
\begin{eqnarray}
E_{O}(A,J)\; &=&\int d^{3}x\left( \frac{1}{4}F_{ij}O^{2}F_{ij}+\frac{1}{2}%
F_{ij}OJ_{ij}+\frac{1}{4}J_{ij}J_{ij}+\frac{1}{2}m^{2}A_{i}A_{i}\right)
\label{erex} \\
&=&\int d^{3}x\left( \frac{1}{2}A_{i}\left( -O^{2}\partial ^{2}+m^{2}\right)
A_{i}+\frac{1}{q}A_{i}\epsilon _{ijk}O\partial _{j}I_{k}+\frac{1}{2q^{2}}%
I_{k}I_{k}\right) \;.  \nonumber
\end{eqnarray}
For the vortex solution we get now

\begin{equation}
A_{i}=-\frac{1}{q}\frac{\epsilon _{ijk}}{\left( -O^{2}\partial
^{2}+m^{2}\right) }O\partial _{j}I_{k}\;,  \label{vreg}
\end{equation}
from which it follows that
\begin{equation}
\frac{E_{O}}{L}=\frac{m^{2}n^{2}}{2q^{2}}\int d^{2}k_{\bot }\frac{1}{k_{\bot
}^{2}\widetilde{O}^{2}(k_{\bot })+m^{2}}\;,  \label{ef}
\end{equation}
where
\begin{equation}
\widetilde{O}(p)=\int d^{3}xe^{-ipx}O(x)\;,  \label{ft}
\end{equation}
is the Fourier transform of the kernel $O(x)$. In the local case, \textit{%
i.e. }$\widetilde{O}(k_{\bot })=1$, we recover the expression $\left(
\mathrm{{\ref{luv}}}\right) $ with its ultraviolet logarithmic divergence.
However, in the case where in the ultraviolet limit $\widetilde{O}(k_{\bot
}) $ behaves as $k_{\bot }^{\alpha }$, $\alpha >0$, the energy per unit
length is finite, no matter how small $\alpha $ is. We see thus that the
requirement of finite energy can be fulfilled by means of the inclusion of
suitable nonlocal gauge invariant terms. These terms are expected to show up
as the result of the functional integration over the radial fluctuations of
the order parameter in eq.$\left( \mathrm{{\ref{pr}}}\right) $.

\section{ Conclusion}

In this paper the bosonization rules $\left( \mathrm{{\ref{u}}}\right) $ and
the dual mapping $\left( \mathrm{{\ref{u1-curr}}}\right) -\left( \mathrm{{%
\ref{udual}}}\right) $ have been applied to analyze some features of
superfluids and superconductors. In the case of superfluids, we have been
able to recover the energy of an ensemble of vortex lines in the $^{4}%
\mathrm{He}$ superfluid by use of the dual mapping $\left(
\mathrm{{\ref {u1-curr}}}\right) -\left(
\mathrm{{\ref{udual}}}\right) $. In particular, we have derived a
kind of bound for the energy configuration which, to some extent,
shares similarity with the Bogomol'nyi bound of superconductors.
When the bound is saturated, the dual mapping applies and one
recovers the energy of an ensemble of vortex lines in terms of a
dual electromagnetic action for charged point-like particles. The
world-lines of these charges correspond to the vortex lines. In
the case of superconductors, after recasting the partition
function in terms of an effective pure gauge theory, the existence
of finite energy vortex solutions with the correct flux
quantization has been established.

\section*{Acknowledgments}

The Conselho Nacional de Desenvolvimento Cient\'{\i }fico e Tecnol\'{o}gico
CNPq-Brazil, the CAPES-Brazil {and the SR2-UERJ are acknowledged for the
financial support. }


\begin{thebibliography}{99}
\bibitem{2b}  S. Coleman, Phys. Rev. \textbf{D11}, 2088 (1975); \newline
S. Mandelstam, Phys. Rev. \textbf{D11}, 3026 (1975).

\bibitem{result1}  E.C. Marino, Phys. Lett. \textbf{B263}, 63 (1991).

\bibitem{result2}  E. Fradkin and F.A. Schaposnik, Phys. Lett \textbf{\ B338}%
, 253 (1994).

\bibitem{result3}  N. Bralic, E. Fradkin, V. Manias and F.A. Schaposnik,
Nucl. Phys. \textbf{B446, }144 (1995);\newline
J.C. Le Guillou, E. Moreno, C. Nu\~{n}ez and F.A. Schaposnik, Phys. Lett.
\textbf{B409, }257 (1997)\textbf{; }\newline
F.A. Schaposnik, \textit{Bosonization in $d>2$ dimensions}, Trends in Theor.
Physics, CERN-Santiago de Compostela-La Plata Meeting, La Plata, April 1997,
hep-th/9705186.

\bibitem{result4}  C.P. Burgess and F.Quevedo, Nucl. Phys. \textbf{B421},
373 (1994).

\bibitem{result5}  R. Banerjee, Phys. Lett. \textbf{B358}, 297 (1995);
\newline
R. Banerjee, Nucl. Phys. \textbf{B465}, 157 (1996);\newline
R. Banerjee and E.C. Marino, Mod. Phys. Lett. \textbf{A14, }593 (1999);

R. Banerjee, C. Wotzasek, Nucl.Phys. \textbf{B527,} 402 (1998)

\bibitem{result6}  D.G. Barci, C.D. Fosco and L.E. Oxman, Phys. Lett.
\textbf{\ B375,} 267 (1996).

\bibitem{result7}  A.~Kovner and P.~S.~Kurzepa, Int.\ J.\ Mod.\ Phys.\
\textbf{A9}, 4669 (1994).

\bibitem{det}  D.G. Barci, V.E.R. Lemes, C. Linhares de Jesus, M.B.D. Silva
Maia Porto, S.P. Sorella, Nucl. Phys. \textbf{B524}, 765 (1998).

\bibitem{hc}  D.G. Barci and L. E. Oxman, Nucl. Phys. \textbf{B580}, 721
(2000).

\bibitem{p}  D.G. Barci, J.F. Medeiros Neto, L.E. Oxman, S.P. Sorella, Nucl.
Phys. \textbf{B600} 721 (2001).

\bibitem{un}  D.G. Barci, L.E. Oxman and S.P. Sorella, Phys. Rev. \textbf{D59%
} (1999) 105012.

\bibitem{bi}  E. Harikumar, A. Khare, M. Sivakumar, P. K. Tripathy, Nucl.
Phys. \textbf{B618} (2001) 570.

\bibitem{dual}  C. D. Fosco, V. E.\ R. Lemes, L. E. Oxman, S.P. Sorella, O.
S. Ventura, Ann. Phys. \textbf{290} (2001) 27.

\bibitem{kl}  H. Kleinert, \textit{Gauge Fields in Condensed Matter,} Vol.I,
World Scientific, 1989.

\bibitem{po}  V.N. Popov, \textit{Functional integrals and collective
excitations,} Cambridge University Press, 1987.

\bibitem{bg}  E. B. Bogomolny, Sov. J. Nucl. Phys. \textbf{24}, 449 (1976).

\bibitem{z}  F.V.Gubarev, L. Stodolsky, V.I. Zakharov, Phys. Rev. Lett.
\textbf{86, }2220 (2001).

\bibitem{an}  D. Antonov,  Int. J. Mod. Phys. \textbf{A14,} 4347 (1999).

N. Agasian, Dmitri Antonov, JHEP \textbf{0106,} 058 (2001).

\bibitem{ant}  M. Kiometzis, H. Kleinert, A. M. J. Schakel, Fortschr. Phys.
\textbf{43}, 697 (1995).

\bibitem{ano}  A. A. Abrikosov, Sov. Phys. JETP \textbf{5}, 1174, (1957);

H. B. Nielsen and P. Olesen, Nucl. Phys. \textbf{B61}, 45 (1973).

\bibitem{bl}  Y. Koma, M. Koma, D. Ebert, H. Toki, JHEP \textbf{0208,} 047
(2002).

\bibitem{os}  L. E. Oxman, S.P. Sorella, Phys. Lett. \textbf{\ B531,} 305
(2002).
\end{thebibliography}
\end{document}